\documentclass[11pt]{article}
\usepackage{geometry}
\usepackage{amsthm}
\usepackage{amsmath}
\usepackage{amsfonts}
\usepackage{graphicx}

\def\sign{\operatorname{sign}}

\def\R{\mathbb R}

\renewcommand{\cos}{\operatorname{cos}}
\renewcommand{\sin}{\operatorname{sin}}

\renewcommand{\Im}{\operatorname{Im}}

\title{Interesting Open Problem Related to Complexity of Computing the Fourier Transform and Group Theory}
\author{Nir Ailon \\ CS, Technion \\\texttt{nailon@cs.technion.ac.il}\thanks{This Ongoing Work is Funded by ERC Grant SpeedInfTradeoff}}

\begin{document}

\maketitle
\def\spectral{{\operatorname{spec}}}
\def\sign{\operatorname{sgn}}
\def\dim{n}
\def\KL{{\operatorname{KL}}}
\def\C{\mathbb C}
\def\R{\mathbb R}
\def\P{\mathcal P}
\def\Q{\mathcal Q}
\def\trace{\operatorname{tr}}
\def\diag{\operatorname{diag}}
\def\rank{\operatorname{rank}}
\def\F{{\mathcal F}}
\def\Id{\operatorname{Id}}
\def\f{{\hat f}}
\def\Z{{\mathbb Z}}
\def\X{{\cal X}}
\def\B{{\cal B}}
\def\Ellips{{\cal E}}
\def\S{{\cal S}}
\def\err{{\operatorname{err}}}
\def\MLAE{{\operatorname{MLAE}}}
\def\N{{\mathcal N}}
\def\tr{\operatorname{tr}}
\def\consts{\mathcal{C}}
\def\ring{\mathcal{R}}
\def\Im{{\operatorname{Im}}}
\def\DFT{\operatorname{DFT}}
\def\conv{\operatorname{CONV}}
\def\poly{\operatorname{poly}}
% ----------------------------------------------------------------

\setcounter{page}{0}
\begin{abstract}
The Fourier Transform is one of the most important linear transformations used in science and engineering.  Cooley and Tukey's Fast Fourier Transform (FFT) from 1964 is a method for computing this transformation in time $O(n\log n)$.   From a lower bound perspective,  relatively little is known.  Ailon shows in 2013 an $\Omega(n\log n)$ bound for computing the normalized Fourier Transform assuming only unitary operations on pairs of coordinates is allowed.   The goal of this document is to describe a natural open problem that arises from this work, which is related to group theory, and in particular to representation theory.
\end{abstract}

\section{Introduction}
The (discrete) normalized Fourier transform is a complex linear mapping sending an input $x\in \C^n$ to $y=Fx\in \C^n$, where $F$ is an $n\times n$ unitary matrix defined by
\begin{equation}\label{dft} F(k,\ell) = n^{-1/2}e^{-i2\pi k\ell/n}\ .
\end{equation}
%The Fast Fourier Transform (FFT) of Cooley and Tukey \cite{CooleyT64} is a method for computing the Fourier transform 
%of a vector $x\in \C^n$
%in time $O(n\log n)$ using a so called linear-algebraic algorithm. 

If $n$ is a power of $2$, then Walsh-Hadamard transform is a real, orthogonal mapping $H$, with the element in position $(k,\ell)$ given by:
\begin{equation}\label{WH} H(k,\ell) = n^{-1/2}(-1)^{\langle k,\ell\rangle}\ ,
\end{equation}
where $\langle k,\ell\rangle$ is the dot-product modulo $2$ of the binary representations of the integers $k-1$ and $\ell-1$.

More genererally, both $F$ and $H$ (resp.) are defined by the characters of underlying abelian groups $\Z/n\Z$ (integers modulo $n$, under addition) and the $\log n$ dimensional binary cube $(\Z/2\Z)^{\log n}$ ($\log n$ dimensional vector space over bits).  Given an input vector $x\in \C^n$, it is possible to compute $Fx$ and $Hx$ (resp.) in time $O(n\log n)$ using the 
Fast Fourier-Transform  \cite{CooleyT64} , or the Walsh-Hadamard transform (resp.).

As for computational lower bounds, it is trivial  that computing both $Fx$ and $Hx$ requires a linear number of steps,
because each coordinate of the output depends on all the input coordinates.

There has not been much prior work on better bounds. We refer the reader to \cite{Ailon13} for a brief history of this line of work,
and concentrate on a recent lower bound.

The work \cite{Ailon13} provides a lower bound of $\Omega(n \log n)$ operations for computing $Fx$ (or $Hx$) given $x$,
assuming that at each step the computer can perform a unitary operation affecting at most $2$ rows. In other words, the
algorithm, running in $m$ steps,  is viewed as a product $$R_m\cdot R_{m-1}\dots R_2\cdot  R_1$$ of matrices $R_t$, each $R_t$ a block-diagonal matrix with $n-2$  blocks equalling $1$, and one block equalling a $2\times 2$ unitary matrix $A_t$:
\begin{equation}\label{22}
R_t = \ \ \ \ \ \ 
\left (
\begin{matrix}
\begin{matrix} 1 & & \\ & \ddots & \\ & & 1 \end{matrix} & \begin{matrix} \phantom{0}^{i_t} \\ \phantom{vdots} \\ \phantom{1}\end{matrix} & &  \begin{matrix} \phantom{0}^{j_t} \\ \phantom{vdots} \\ \phantom{1}\end{matrix}&  \\
\!\!\!\!\!\!\!\!\!\!\!\!\!\!\!\!\!\!\!\!\!\!\!\!\!\!\!\!\!\!\!\!\!\!\!\! i_t& A_t(1,1) & & A_t(1,2)& \\
& & \begin{matrix} 1 & & \\ & \ddots & \\ & & 1 \end{matrix}   & & \\
\!\!\!\!\!\!\!\!\!\!\!\!\!\!\!\!\!\!\!\!\!\!\!\!\!\!\!\!\!\!\!\!\!\!\!\! j_t& A_t(2,1) & & A_t(2,2) & \\
& & & & \begin{matrix} 1 & & \\ & \ddots & \\ & & 1 \end{matrix} 
 \end{matrix}
 \right )\ .
 \end{equation}
 
 The justification for this model of computation is threefold:
 \begin{itemize}
 \item  Similarly to matrices of the form (\ref{22}), any basic operation of a modern computer (e.g., addition of numbers) 
 acts on only a fixed number of inputs. 
 \item The Fast Fourier-Transform, as well as the Walsh-Hadamard transform, operate in this model, and
 \item The set of matrices of the form (\ref{22}) generate the group of unitary matrices.
 \end{itemize}
 Thus the question of computational complexity of of the Fourier transform becomes that of computing distances between 
elements of a group, namely the unitary group, with respect to a set of generators that is computationally simple.

 Obtaining the lower bound of $\Omega(n\log n)$ in \cite{Ailon13} is done by defining a potential function $\Phi$ for unitary
 matrices, as follows:
 $$
 \Phi(U) = -\sum_{i,j} |U(i,j)|^2\log |U(i,j)|^2
 \ .
 $$
 
  With this potential function, one shows that:
  \begin{enumerate}
  \item[(a)]  $\Phi(\Id)=0$
  \item[(b)] $\Phi(F) = \Phi(H) = \Omega(n\log n)$
  \item[(c)] $|\Phi(M_t) - \Phi(M_{t-1})| \leq 2$, where $M_t = R_t\cdot R_{t-1}\dots R_1$ is the \emph{state}
  of the algorithm after $t$ steps.
  \end{enumerate}
  Indeed, if the potentail $\Phi$ grows from $0$ to $\Omega(n\log n)$ changing (in absolute value) by no more  than $2$ at each step, then the number of steps must be $\Omega(n\log n)$.  Showing $(c)$ is done using two observations. The first is   that $M_t$ defers from $M_{t-1}$ in at most $2$ rows $i_t$ and $j_t$, and that  for each column $k$, due to unitarity of
  $A_t$, $$|M_t(i_t, k)|^2 + |M_t(j_t, k)|^2 = |M_{t-1}(i_t, k)|^2 + |M_{t-1}(j_t, k)|^2\ .$$
The next observation is that  any $4$ numbers $x,y,z,w$ satisfying $x^2+y^2  = z^2+w^2 =: r^2$, also satisfy 
  $$ |(x^2\log x^2 + y^2 \log y^2) -  (z^2\log z^2 + w^2 \log w^2) | \leq r^2\ .$$
Combining the observations, we conclude that the total change in the potential function can be at most
$$ \sum_k (|M_t(i_t, k)|^2 + |M_t(j_t, k)|^2)  = 2\ .$$

\section{An Interesting  Problem}

The advantage of the method just described is that it reduces a computational problem to that of computing distance between 
two elements of a group, with respect to a chosen set of generators of the group.  We now define a more general problem within the same group theoretical setting.

\noindent
Consider the $2n\times 2n$ matrix $G$ defined as
$$ G = \left ( \begin{matrix} 0 & F \\ -F^* & 0 \end{matrix} \right )\ .$$
(One may replace $F$ with $H$, but we work with $F$ henceforth.)
The matrix $G$ is skew-Hermitian.  Let $\Id$ denote the $2n\times 2n$ identity matrix, and finally define for a real angle $\alpha$ the following matrix:
$$ X_\alpha = (\cos \alpha) \Id + (\sin \alpha) G\ .$$

\noindent
It is easy to verify that $X_\alpha$ is unitary for all $\alpha$.  It is also easy to verify that
\begin{equation}\label{030303}
X_{\alpha'} X_\alpha = X_{\alpha+\alpha'}\ .
\end{equation}
 Also, using the potential function $\Phi$ defined above, we see that
$$ \Phi(X_\alpha) = \Theta(\alpha^2 n \log n)\ .$$
Hence, using the argument as above, the number of steps required to compute $X_\alpha$ must be at least $\Omega(\alpha^2 n\log n)$. However,  it is unreasonable that it should be possible to compute $X_\alpha$ faster than the time it takes to compute $F$, by a factor  of $1/\alpha^2$.  Indeed, given an input $x\in \C^n$, we could simply embed it as $\tilde x\in \C^{2n}$ by padding with $n$ $0$'s,
then compute $\tilde y = X_\alpha \tilde x$ and then retrieve $y=Fx$ from $\tilde y$ and $\tilde x$ by a simple arithmetic manipulation.  Hence, we conjecture that the number of steps required to compute $X_\alpha$ should be not much smaller than $\Omega(n\log n)$.
\footnote{The author conjectures $\Omega((n\log n)/\log(1/\alpha))$ to be the correct bound.}

\subsection{A slight improvement: Lower bound of $\Omega (\alpha n\log n)$.}
It is possible to get a better bound than $\Omega(\alpha^2 n\log n)$, as follows.
Instead of starting the computation at state $\Id$ and finishing at $X_\alpha$, we can opportunistically choose a starting
point $M$ (and finish at $X_\alpha M$).

If we choose the state $M = X_{\pi/4 - \alpha/2}$ then it is trivial to verify that the computation ends at state $X_\alpha M$, which equals   $X_{\pi/4 + \alpha/2}$ by (\ref{030303}).
We then observe that
$$ \Phi(X_{\pi/4 - \alpha/2}) = \Theta\left (\sin^2(\pi/4 - \alpha/2)\cdot n\log n\right )\ \ \ \ \ \Phi(X_{\pi/4+\alpha/2}) = \Theta\left (\sin^2(\pi/4+\alpha/2)\cdot n\log n\right )\ ,$$
and hence,
\begin{eqnarray*}
 \left | \Phi(X_{\pi/4 + \alpha/2})  - \Phi(X_{\pi/4 - \alpha/2})  \right | &=& \Omega\left ((\sin^2(\pi/4 + \alpha/2) - \sin^2(\pi/4 - \alpha/2))n\log n\right ) \\ 
 &=& \Omega\left ( \alpha n \log n \right )\ .
 \end{eqnarray*}
 \subsection{Stronger improvements?}
 Is it possible to get a stronger lower bound than $\alpha n \log n$?  
 One approach for solving this problem might be using group representation theory.
 If $\Psi : U(n) \mapsto U(n')$ is any unitary  representation of $U(n)$, then we could define a new potential function
 $\Phi \circ \Psi$ on $U(n)$, and use it to obtain possibly better lower bounds.
 
An interesting representation is related to determinants.
We let the order $k$ determinant representation of a unitary matrix $U$ be the matrix $\Psi_{det}^k(U)$ of shape ${n\choose k}\times {n\choose k}$, defined by
$$(\Psi_{det}^k(U))_{I, J} = \det{U_{I,J}}\ ,$$
where $I,J$ are subsets of size exactly $k$ of $[n]$, $U_{I,J}$ is the $k$-by-$k$ submatrix defined by row set $I$ and column set $J$.
The fact that $\Psi_{det}^k(U)$ is a unitary matrix coming from a group representation is non-trivial, and we refer the reader to
resources on representation theory for more details.

So far I have not been able to make progress on the problem using this (quite natural) representation, but I am not convinced that this
direction is futile either.

\subsection{Important Note: Even the case $\alpha = \pi/4$ is Interesting}
Note that although the main problem proposed here is to understand the asymptotic behviour of the complexity of $X_\alpha$, as $\alpha$ tends to $0$, even the case of finding a lower bound for computation of $X_{\pi/4}$ is not trivial, in the sense that it is not
clear how (and whether it is at all possible) to get a bound better than $\frac 1{2\sqrt 2} n \log_2 n$, which is the best
possible using the ``vanilla'' entropy function $\Phi$.
\bibliography{low_bound_fft}
 \end{document}